# Films of Mn$_{12}$-acetate by Pulsed Laser Evaporation

V. Meenakshi, W. Teizer[*], D. G. Naugle, H. Zhao[a] and K. R. Dunbar[a]

Department of Physics, Texas A&M University, College Station, TX 77843

[a]Department of Chemistry

**Abstract**

Films of the molecular nanomagnet, Mn$_{12}$-acetate, have been deposited using pulsed laser deposition and a novel variant, matrix assisted pulsed laser evaporation. The films have been characterized by X-ray photoelectron spectroscopy, mass spectrometry and magnetic hysteresis. The results indicate that an increase in laser energy and/or pulse frequency leads to fragmentation of Mn$_{12}$, whereas its chemical and magnetic integrity is preserved at low laser energy (200 mJ). This technique allows the fabrication of patterned thin film systems of molecular nanomagnets for fundamental and applied experiments.

## 1. Introduction

As new molecular magnetic materials of fundamental and technological importance are discovered,[1] there is a strong impetus to process them in reproducible thin film form. Potential technological applications (e.g. in magnetic storage devices) of this new class of materials require complete control of the organization of these molecules on a substrate [2]. Well-established film growth techniques that suffice for the deposition of simple materials fail to meet the demands required by these chemically fragile molecular magnets. Therefore, despite possessing a range of interesting properties[3,4], (such as high magnetic anisotropy, large magnetic hysteresis of purely molecular origin, etc.) and exhibiting quantum tunneling of magnetization, deposition challenges have precluded workers in the field from producing well-controlled thin film structures. Hence, the aim of this paper is to address the problem of depositing reproducible films of molecular nanomagnets without destroying the magnetic properties.

---

[*] Corresponding author.
  E-mail address: teizer@tamu.edu (W. Teizer)

The molecule, $Mn_{12}O_{12}(CH_3COO)_{16}(H_2O)_4 \cdot 2CH_3COOH \cdot 4H_2O$, is one of the most thoroughly studied [5-10] single molecule magnets (SMM), commonly referred to as $Mn_{12}$, and consists of twelve Mn ions in $Mn^{III}$ and $Mn^{IV}$ oxidation states. These are molecules in which the magnetic domain size can be reduced to a single molecule below a certain temperature, known as the blocking temperature. Until recently, all studies on $Mn_{12}$ have been limited to single crystal or polycrystalline 3-d samples. The first attempt to organizing these molecules as multilayers was undertaken by using the Langmuir-Blodgett (LB) technique to produce magnetic monolayers of $Mn_{12}$ clusters [11]. SMM complexes other than $Mn_{12}$-acetate [12] have also been placed on surfaces. Another interesting method involves designing derivatives of $Mn_{12}$ clusters with suitable surface-binding functionalities, that adhere to specific surfaces (e.g. Au) [13]. Among these various approaches, most involve a functional modification of the starting complex.

The main reason for this relatively small amount of film work is the low decomposition temperature (~ $65^{\circ}C$) of $Mn_{12}$, which prevents the use of conventional thin film deposition techniques, such as, thermal evaporation, sputtering, e-beam evaporation, etc. For $Mn_{12}$ molecules, as revealed by Thermo Gravimetric Analysis, the loss of $H_2O$ and acetic acid molecules of solvation begins above T ~ $35^{\circ}C$ and continues until T ~ $94.2^{\circ}C$ [14]. The core of the $Mn_{12}$ molecules fragment above T ~ $195^{\circ}C$ leading to the complete decomposition of the compound upon further heating. The useful magnetic properties of molecular $Mn_{12}$ are lost in the process.

Here we report on depositing $Mn_{12}$ films using a novel variation of the conventional laser ablation: matrix assisted pulsed laser evaporation (MAPLE). This technique has been previously used for depositing films of easily decomposable polymers and organic compounds [15-16]. It allows a flexible choice of film thickness and substrate material. For comparison, $Mn_{12}$ films were also deposited by conventional pulsed laser deposition (PLD). Films were deposited with different values of laser energy (200 mJ - 450 mJ) as well as laser pulse frequency (1 - 3 Hz). The $Mn_{12}$ films deposited using MAPLE and PLD, along with the as-produced $Mn_{12}$ have been characterized using mass spectrometry (MS), X-ray photoelectron spectroscopy (XPS) and magnetic hysteresis measurements. The central result of the present study is that, for the first time, continuous films of $Mn_{12}$ have been deposited using MAPLE and PLD. The MAPLE films exhibit superior quality as compared to those produced by PLD. We also find that higher laser energy (above 250 mJ) and pulse frequency (3 Hz) result in fragmentation of $Mn_{12}$, thereby establishing a threshold below which quality $Mn_{12}$ films can be deposited.

## 2. Experimental Details

*2.1 Pulsed Laser Deposition (PLD) of $Mn_{12}$*

In PLD, the target consists of a pellet (made by compressing $Mn_{12}$ powder to a pressure of 20000 psi) mounted in a stainless steel holder. A Lambda Physik Excimer laser operating with a KrF gas mixture (10 ns pulse width at 248 nm) firing at rates between 1-3 Hz was used. The laser was focused onto a 0.25 $cm^2$ area and the laser fluence at the target was varied between 200 - 450 mJ. The target was rotated in order to avoid excessive heating of a single spot on its surface. The target to substrate distance was 5 cm and the substrates used were either fire polished glass or $Si/SiO_X$ (X = 1,2, native oxide). The substrates were kept at room temperature. The chamber pressure prior to deposition was ~ $10^{-3}$ Torr and during deposition it varied between $10^{-3}$ Torr ~$10^{-1}$ Torr. Typical deposition times used for PLD were ~ 45 - 60 minutes and the thickness of the films obtained was ~ 800 - 1000 Å.

*2.2 Matrix Assisted Pulsed Laser Evaporation (MAPLE)*

MAPLE (for schematic, see Figs. 1 and 2 in Ref. 15) is a variation of the conventional pulsed laser evaporation. It is a more gentle mechanism for transferring compounds from the condensed phase to the vapor phase. The key to this technique is the use of a frozen target consisting of a dilute solution of the material of interest in a relatively volatile solvent (like water, methanol, chloroform, etc) matrix. The laser interaction primarily occurs with the solvent, and the material of interest (solute) remains intact. The solute concentration is intentionally kept low so that the incident laser energy is mostly absorbed by the solvent molecules rather than by the solute molecules. The photo-thermal process initiated at the molecular level by the laser pulse causes the solvent to vaporize and the collective action of multiple collisions of the evaporating solvent with the embedded solute results in the soft desorption of the latter, with little or no damage to its structure and functionality. When a substrate is positioned in the path of the plume, a continuous film is formed from the evaporated solute, while the lower molecular weight solvent is rapidly removed by continuous evacuation. The substrates as well as deposition conditions and parameters (laser energy, pulse frequency) were identical to those used for PLD. Deposition time used for MAPLE varied between 40 - 80 minutes and typical thickness (by ellipsometry) of the films was ~ 50 - 300 Å.

*2.3 MAPLE and the thermal stability of $Mn_{12}$*

Due to the low decomposition temperature of $Mn_{12}$, an important issue in using a laser deposition method for $Mn_{12}$ films is the ability of these thermally unstable molecules to withstand the energetic environment created by a laser pulse. As discussed in detail by Vertes [17], *molecules can survive volatilization if their liberation by disintegration precedes their destruction by fragmentation*. This can be achieved by a faster heating rate, which is more likely to promote vaporization over decomposition. In this way, the adhesive/cohesive bonds between the solvent and the molecule of interest break before sufficient energy could be transferred to heat up the molecule of interest. Cold molecules can then be liberated, without damage to their structure. This is the principle behind MAPLE.

In order to deposit films of $Mn_{12}$ using MAPLE, it is therefore crucial to identify a suitable solvent with appropriate volatility, $Mn_{12}$ solubility and laser absorption. The solubility of $Mn_{12}$ in common organic solvents (e.g. acetone) is low and, in many cases, $Mn_{12}$ decomposes after approximately an hour. It is convenient if the solvent also possesses a high freezing temperature, as the target has to be maintained in a frozen state during ablation. All the above requirements were met by tert-butyl alcohol, which has a freezing temperature of 26°C. A liquid nitrogen cooling assembly was used to maintain the target frozen at all times during the laser ablation process. For MAPLE, it is sufficient if the concentration of $Mn_{12}$ in the solvent is moderate. The concentration of the solution was in the range $10^{-4}$ M - 5 x $10^{-3}$ M (Higher concentration results in some undissolved $Mn_{12}$). To ensure that $Mn_{12}$ did not interact with the solvent, after dissolving a considerable quantity of as-produced $Mn_{12}$ in the solvent, the powder was re-crystallized and X-ray diffraction was performed on the resultant material confirming that the $Mn_{12}$ after recrystallization was identical to that of the starting material.

*2.4 Details of Characterization*

The surface morphology of the deposited films was examined using an Atomic Force Microscope (AFM). The films were observed to be continuous over the area scanned, 0.5μ x 0.5μ (Fig. 1). The topographical view reveals interconnected spherical clusters of varying sizes. From the height analysis, the root-mean-square roughness was estimated to be ~ 13 Å ($Mn_{12}$ molecular diameter is ~ 17 Å). The AFM image of bare Si substrate showed no surface corrugations and has a roughness of ≤ 5 Å.

The films and the bulk $Mn_{12}$ starting material were characterized by several techniques. Electrospray ionization mass spectra were recorded on a PE Sciex API Qstar Pulsar. Core level XPS measurements were performed using a Kratos Axis HSi 165a Ultra X-ray photoelectron spectrometer at room temperature by using Al (hν = 1486.6 eV) and Mg (hν = 1253.6 eV) $K_α$ radiation from a twin anode. The chamber pressure was maintained at $10^{-8}$ Torr, and the energy resolution was better than 0.5 eV. Hysteresis measurements were performed using a Quantum Design SQUID magnetometer MPMS-XL at 1.8K and 15 K. For magnetic measurements, polycrystalline as-prepared $Mn_{12}$ powder was used, and, in the case of deposited PLD films, the powder was removed from the substrate after film deposition.

## 3. Results and discussion

### 3.1 Mass Spectrometry

Electrospray ionization mass spectrometry (MS) provides an accurate determination of the molecular mass, and hence is useful for determining the integrity of the $Mn_{12}$ molecules after ablation. For the MS studies, powder was removed from the substrate after PLD, was then dissolved in acetonitrile and the instrument was operated in the negative ion mode: The $Mn_{12}$ clusters are neutral, but they undergo reduction in the MS studies to form multiply charged species.

The results for two PLD films deposited at ~ 200 mJ laser energy but at two different pulse frequencies 1 and 3 Hz are presented in Figs. 2(a) and 2(b), respectively. Note that at the pulse frequency of 3 Hz, the average power deposited is three times greater than at 1 Hz. The highlighted boxes in Fig. 2(a) denote the three peaks pertaining to the original $Mn_{12}$ sample in acetonitrile. The presence of a molecular peak at m/z = 1795 corresponds [18] to $[Mn_{12}O_{12}(CH_3COO)_{16}]^{1-}$. Fragmentation of this species by successive loss of one acetate ligand [18] gives rise to the peak at m/z = 1736. The presence of the doubly reduced species $[Mn_{12}O_{12}(CH_3COO)_{16}]^{2-}$ is clearly evident from the presence of a molecular peak [18] at m/z = 897. It can be seen that even for the lowest pulse frequency (lowest average power) used during PLD, additional peaks are observed that are not present in the MS of the as-produced bulk sample. Furthermore, in Fig. 2(b), an increase in laser pulse frequency results in the appearance of more MS peaks in the mass range 800-1300, indicating fragmentation of $Mn_{12}$ clusters in the deposited film. The greater fragmentation of $Mn_{12}$ clusters at 3 Hz causes the molecular peak intensities corresponding to the parent molecule $Mn_{12}O_{12}(CH_3COO)_{16}$ to become comparable to the intensity of the fragment peaks. We also deposited films at laser energies greater than 200 mJ and observed complete fragmentation above 400 mJ (not shown).

The MS analysis reveals that, even with the use of a low laser energy (~ 200 mJ), the deposited film contains small amounts of fragments of $Mn_{12}$. Therefore, we employed a more gentle transfer mechanism – MAPLE - for depositing the $Mn_{12}$ films. The films deposited using MAPLE were very thin (~ 50 - 300 Å) and hence it was impossible to extract sufficient powder from the substrate for the MS analysis. Therefore, we characterized these films using X-ray photoelectron spectroscopy, which allows a comparison of the electronic structure of the films deposited from PLD, MAPLE and the as-produced $Mn_{12}$.

*3.2 Magnetic Hysteresis*

Prior to shifting the focus to MAPLE deposited films, we present results from the magnetic measurements on the PLD films produced at ~200 mJ laser energy, 1 Hz pulse frequency to demonstrate that the magnetic properties of some $Mn_{12}$ molecules are preserved. Data from experiments initially performed on films deposited on Si/glass substrates had a poor signal-to-noise and hence it was decided to deposit thick films and to remove the powder from the substrate for the measurement. Consequently, the sample used for the measurement contains a collection of randomly oriented molecules. Films deposited by MAPLE on Si/glass substrates were too thin, such that their magnetic measurements were dominated by the substrate contribution. In this case, it was impossible to remove sufficient material from the substrate (as was done for PLD films) because the films were thin. Data for them is not reported here [19].

The hysteresis loops obtained at 1.8 K and 15 K with powder from the PLD films are shown in Fig. 3. The inset expands the low field behavior. Both at 1.8 K and 15 K, the hysteresis shows an anomalous "pinched" shape near zero field. Ordinary hysteresis loops (i.e., those of oriented $Mn_{12}$) are widest at low fields (near zero) and exhibit an inflection point at the coercive field. A "pinched" hysteresis loop implies that the magnetic relaxation is greater in zero field than it is in small finite magnetic fields. A plausible reason for this observation is that the material contains a random distribution of magnetic particle sizes, and the resonant condition for tunneling can occur only at zero field. It is known that powdered samples of $Mn_{12}$ that are not oriented during preparation show smooth hysteresis loops without any steps and are "pinched" near zero field [20,21]. This is also seen for the LB films [11]. Since the PLD films also contained some fragments (as shown by MS), another possibility may be the presence of more than one phase in the sample [22]. The additional phase probably may be due to the removal of few acetate ligands in the process of PLD. We see evidence for this by comparing the MS of crystalline $Mn_{12}$ and that of the films. Specifically, the ratio of size of the peaks at 1736 (corresponding to the loss of one acetate ligand) and 1795 (corresponding to the original $Mn_{12}$ complex) is increased in PLD films as compared to crystalline $Mn_{12}$. Also, in contrast with the behavior seen in $Mn_{12}$ crystals, the magnetization at 6T is far from saturation, as may be expected for a random distribution of particles.

*3.3 X-ray photoelectron spectroscopy (XPS)*

The XPS technique provides direct information about the electronic structure of the films. Since this technique probes the surface, it is suitable for very thin films. In the present study, XPS allows for a comparison of the crystalline $Mn_{12}$ sample, as well as films deposited using PLD and MAPLE on $Si/SiO_X$. Experiments were also performed on films deposited on fire-polished glass. Core levels due to Mn, C and O ions were observed. Peaks in a core level spectrum reflect the chemical environments of crystallographically different atoms. The existence of multiple peaks in each of the core level spectra signifies the existence of different Mn, C and O sites in $Mn_{12}$ [23]. In all the samples, no peaks from the substrate were observed in the spectra, although a variety of regions of the film on the substrate were scanned. This indicates that the deposited films exhibit continuous coverage.

Figure 4 shows the Mn 2p core level spectrum which has two peaks at ~ 652.5 eV and 640.5 eV corresponding to $2p_{1/2}$ and $2p_{3/2}$ respectively. The plot contains the spectra obtained from crystalline $Mn_{12}$ samples, MAPLE films deposited at laser energies 200 mJ, 350 mJ and a PLD film deposited at a laser energy of 200 mJ. The peak position for the MAPLE film with lowest energy (200 mJ) is identical to that of as-produced $Mn_{12}$ within experimental resolution. As the laser energy is increased to 350 mJ in MAPLE, an obvious shift in the peak position from as-produced $Mn_{12}$ is seen and this shift progressively increases with increased laser energy. For PLD films, even with a laser energy of 200 mJ, the peak position is shifted very clearly to higher binding energy, even more than for the MAPLE films deposited at 350 mJ. The upshift in peak position for the PLD films concurs with the MS results, where fragments of lower molecular mass (Fig. 2) are found in PLD films, in addition to as-produced $Mn_{12}$. This may imply that the binding energy for Mn in a smaller cluster is greater than that in a bigger cluster and hence an upshift in binding energy is seen for the fragmented film. Most importantly, XPS indicates that films from MAPLE are of superior quality than those from PLD, i.e. the $Mn_{12}$ cluster integrity is better preserved during MAPLE at low laser energies. We further confirmed that the shift in peak position with increased laser energy as well as with larger pulse frequency is indeed intrinsic by measuring the secondary electron emission peaks and evaluating the Auger parameter from the difference in peak position of the core and secondary electron emission. Due to the fact that $Mn_{12}$ is an insulator, any peak shift introduced by improper charge neutralization is accounted for in this technique. Similar to Mn 2p peaks, a comparison of O 1s peak positions (from O atoms bridging Mn in the core and those in water and acetic acid) confirm again that MAPLE films (deposited at 200 mJ) match well with the as-produced $Mn_{12}$, unlike PLD films for which changes in peak positions are clearly evident (not shown).

## 4. Conclusions

In summary, we have succeeded in the deposition of continuous thin films of SMM $Mn_{12}$ using the MAPLE and PLD technique. This is a reliable and simple method to organize $Mn_{12}$ as thin films. The MAPLE approach relies on the laser interaction occurring primarily with the solvent resulting in better films with the $Mn_{12}$ molecules intact. Also, laser deposition offers flexibility for controlling film properties (e.g. thickness) and for choice of the substrate, depending on the requirement. XPS shows that MAPLE films are of superior quality as compared to those from PLD. By keeping the laser energy and pulse frequency low, it is possible to improve the quality of films from PLD to retain their molecular magnetism. These results open a path for the fabrication of arbitrarily structured thin film systems of molecular nanomagnets which allow several fundamental and applied experiments.

**Acknowledgements**

We thank C. Berlinguette, V. Pokrovsky and N. Sinitsyn for helpful discussions. This research was supported by the Texas Advanced Research Program (010366-0038-2001) and the Robert A. Welch Foundation. DGN acknowledges the Robert A. Welch Foundation Grant (A.0514). KRD and DGN gratefully acknowledge the National Science Foundation support from Nanoscale Science and Engineering (NIRT) Grant (DMR-0103455), the Telecommunications and Informatics Task Force (TITF 2001-3) at Texas A&M University, and the National Science Foundation for an equipment grant to purchase a SQUID magnetometer (NSF-9974899). Use of the TAMU/CIMS Materials Characterization Facility and discussions with Dr. W. Lackowski and Ms. Y. Vasilyeva are also acknowledged.

**References**

[1] J. S. Miller and A. J. Epstein, MRS Bulletin 25 (2000) 21. A. J. Epstein, MRS Bulletin 25 (2000) 33. G. Christou, D. Gatteschi, D. N. Hendrickson and R. Sessoli, MRS Bulletin 25 (2000) 66.

[2]  D. L. Leslie-Pelecky and R. D. Reike, Chem. Mater. 8 (1996) 1770.

[3]  B. Schwarzschild, Phys. Today 50 (1997) 17.

[4] D. Gatteschi, A. Caneschi, L. Pardi and R. Sessoli, Science 265 (1994) 1054.

[5] R. Sessoli, H. L. Tsai, A. R. Schake, S. Wang, J. B. Vincent, K. Folting, D. Gatteschi, G. Christou and D. N. Hendrickson, J. Am. Chem. Soc. 115 (1993) 1804.

[6] R. Sessoli, D. Gatteschi, A. Caneschi and M. A. Novak, Nature 365 (1993) 141.

[7] W. Wernsdorfer, in : I. Prigogine and S. A. Rice (Eds.), Advances in Chemical Physics, Wiley, New York, vol. 118, 2001, pp. .

[8] A. Caneschi, D. Gatteschi, C. Sangregorio, R. Sessoli, L. Sorace, A. Cornia, M. A. Novak, C. Paulsen, W, Wernsdorfer, J. Magn. Magn. Mater. 200 (1999) 182.

[9] B. Barbara, L. Thomas, F. Lionti, I. Chiorescu and A. Sulpice, J. Magn. Magn. Mater. 200 (1999) 167.

[10] J. M. North, L. J. van de Burgt and N. S. Dalal, Solid State Commun. 123 (2002) 75.

[11] M. Clemente León, H. Soyer, E. Coronado, C. Mingotaud, C. J. Gómez-García and P. Delhaès, Angew. Chem. Int. Ed. 37 (1998) 2842.

[12] Daniel Ruiz-Molina, Marta Mas-Torrent, Jordi Gómez, Ana I. Balana, Neus Domingo, Javier Tejada, María T. Martínez, Concepció Rovira and Jaume Veciana, Adv. Mater. 15 (2003) 42. Massimiliano Cavallini, Fabio Biscarini, Jordi Gomez-Segura, Daniel Ruiz, and Jaume Veciana, Nano Lett, 3 (2003) 1527.

[13] A. Cornia, A. C. Fabretti, M. Pacchioni, L. Zobbi, D. Bonacchi, A. Caneschi, D. Gatteschi, R. Biangi, U. D. Pennino, V. D. Renzi, L. Gurevich and H. S. J. Van der Zant, Angew. Chem. Int. Ed. 42 (2003) 1645.

[14] T. Lis, Acta Cryst. B 36 (1980) 2042.

[15] A. Pique, R. A. McGill, D. B. Chrisney, D. Leonhardt, T. E. Mslna, B. J. Spargo, J. H. Callahan, R. W. Vachet, R. Chung, M. A. Bucaro,Thin Solid Films 355-356 (1999) 536.

[16] A. Pique, P. Wu, B. R. Ringeisen, D. M. Bubb, J. S. Melinger, R. A. McGill, D. B. Chrisey, Appl. Sur. Sci. 186 (2002) 408.

[17] A. Vertes, in : K. G. Standing and W. Ens (Eds.), Methods and Mechanisms for Producing Ions from Large Molecules, Plenum Press, New York, 1998, pp. 275-286.

[18] E. Coronado, M. Feliz, A. Forment-Aliaga, C. J. Gomez-Garcia, R. Llusar and F. M. Romero, Inorg. Chem. 40 (2001) 6084.

[19] Thicker MAPLE films (can be obtained by increasing the $Mn_{12}$ concentration in solvent and also the ablation time) may allow this measurement in the future.

[20] J. M. Hernández, X. X. Zhang, F. Luis, J. Tejada, J. R. Friedman, M. P. Sarachik and R. Ziolo, Phys. Rev. B 55 (1997) 5858.

[21] J. R. Friedman, U. Voskoboynik and M. P. Sarachik, Phys. Rev. B 56 (1997) 10793.

[22] K. Takeda, K. Awaga and T. Inabe, Phys. Rev. B 57 (1998) R11062.

[23] J. S. Kang, J. H. Kim, Y. J. Kim, W. S. Jeon, D. Y. Jung, S. W. Han, J. H. Kim, K. J. Kim and B. S. Kim, J. Korean Phys. Soc. 40 (2002) L402.

**Figure Captions**

Fig. 1. AFM (3-D view, topographical view and height profile, respectively) of MAPLE deposited $Mn_{12}$ film

Fig. 2. ESI-MS of PLD $Mn_{12}$ films deposited at laser pulse frequency (a) 1Hz and (b) 3Hz. Highligted boxes in (a) denote the position of peaks for an original crystalline sample of $Mn_{12}$. The corresponding molecular species are discussed in the text

Fig. 3. Hysteresis of PLD films at 1.8 K and 15 K and the inset shows the low field behavior

Fig. 4. Mn 2p core level spectra of as-produced $Mn_{12}$, films from MAPLE and PLD

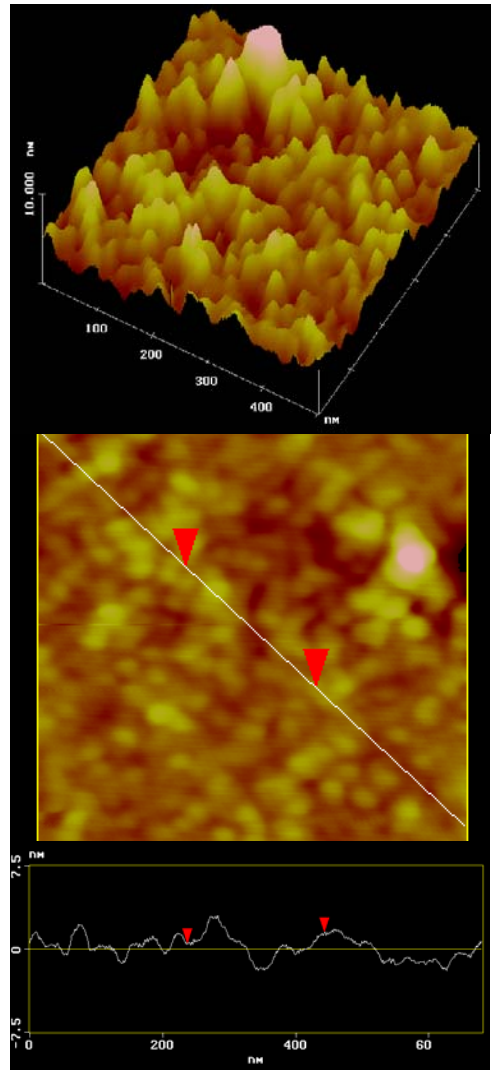

Figure 1

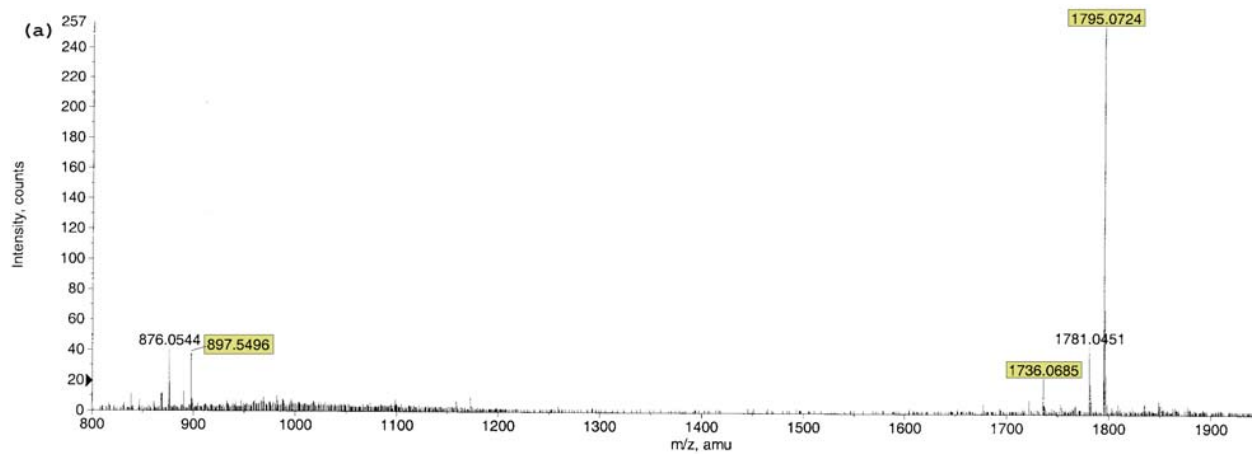

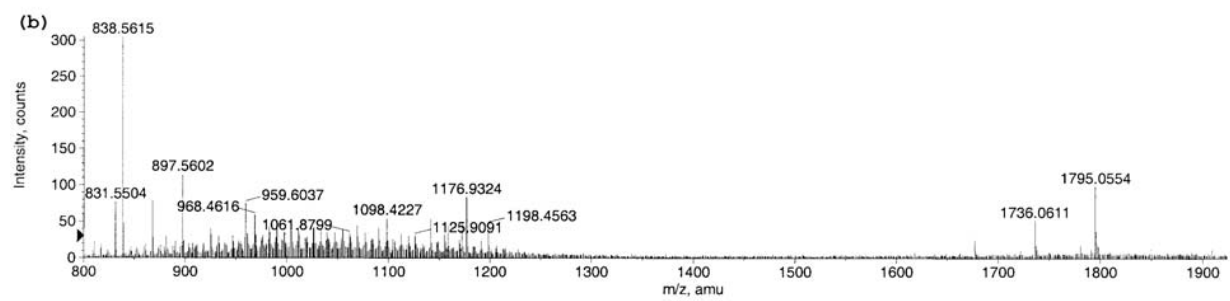

Figure 2

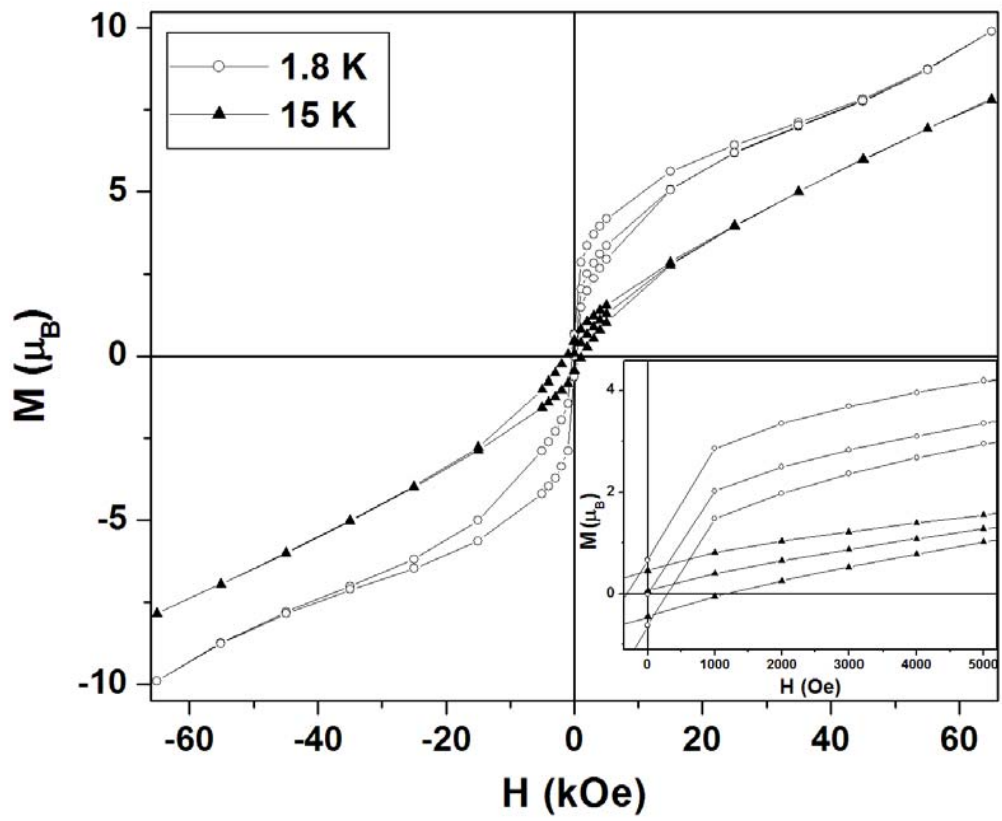

Figure 3

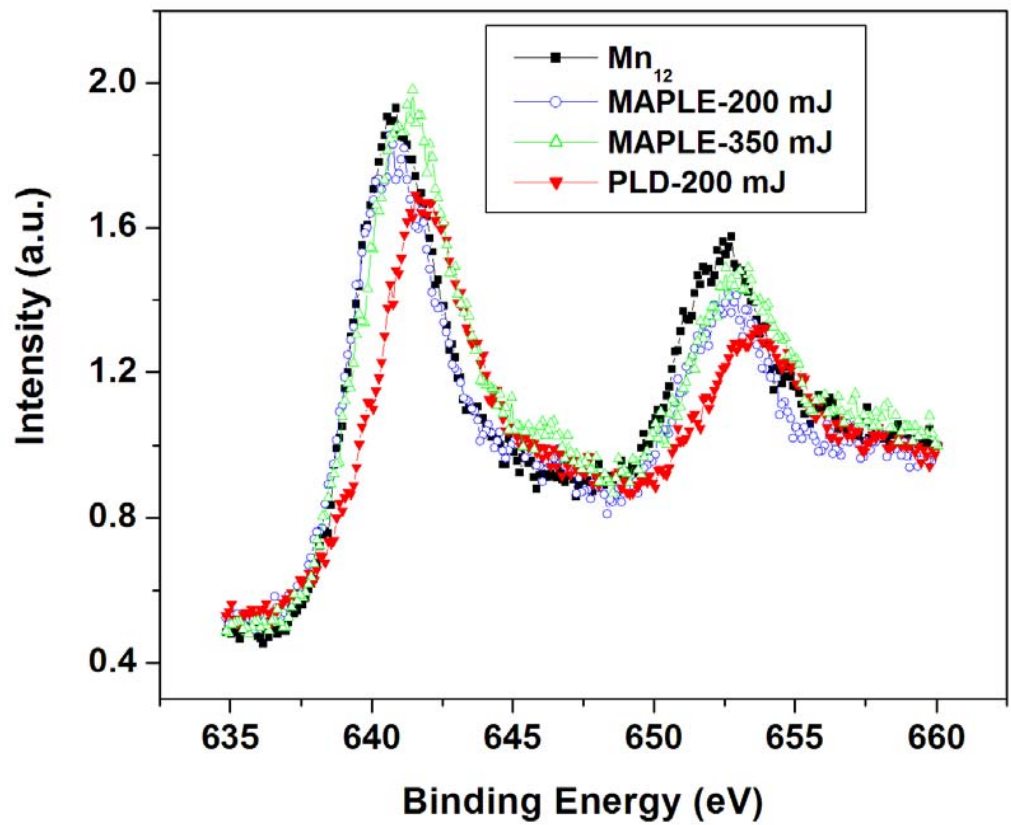

Figure 4